\documentstyle[12pt]{article}
\begin{document}
\tolerance=5000
\def\pp{{\, \mid \hskip -1.5mm =}}
\def\cL{{\cal L}}
\def\be{\begin{equation}}
\def\ee{\end{equation}}
\def\bea{\begin{eqnarray}}
\def\eea{\end{eqnarray}}
\def\tr{{\rm tr}\, }
\def\nn{\nonumber \\}
\def\e{{\rm e}}
\def\D{{D \hskip -3mm /\,}}

\def\SEH{S_{\rm EH}}
\def\SGH{S_{\rm GH}}
\def\AdS5{{{\rm AdS}_5}}
\def\S4{{{\rm S}_4}}
\def\gfv{{g_{(5)}}}
\def\gfr{{g_{(4)}}}
\def\SC{{S_{\rm C}}}
\def\RH{{R_{\rm H}}}


\  \hfill 
\begin{minipage}{3.5cm}
NDA-FP-76 \\
June 2000 \\
\end{minipage}

\vfill

\begin{center}
{\large\bf Bulk versus boundary (gravitational Casimir) effects 
in quantum creation of inflationary brane-world Universe }

\vfill

{\sc Shin'ichi NOJIRI}\footnote{email: nojiri@cc.nda.ac.jp}, 
{\sc Sergei D. ODINTSOV}$^{\spadesuit}$\footnote{On leave 
from Tomsk Pedagogical University, 634041 Tomsk, RUSSIA
email: odintsov@ifug5.ugto.mx, odintsov@mail.tomsknet.ru},

{\sc  and Sergio ZERBINI}$^\diamondsuit$\footnote{
e-mail: zerbini@science.unitn.it }

\vfill

{\sl Department of Applied Physics \\
National Defence Academy, 
Hashirimizu Yokosuka 239, JAPAN}

\vfill

{\sl $\spadesuit$ 
Instituto de Fisica de la Universidad de 
Guanajuato \\
Apdo.Postal E-143, 37150 Leon, Gto., MEXICO} 

\vfill

{\sl $\diamondsuit$ Department of  Physics and Gruppo 
Collegato INFN\\
University of Trento 38050-Trento, ITALY}

\vfill

{\bf ABSTRACT}

\end{center}

The role of bulk matter quantum effects (via the corresponding 
effective potential discussed on the example of conformal scalar) 
and of boundary matter quantum effects (via the conformal anomaly
induced effective action) is considered in brane-world cosmology.
Scenario is used where brane tension is not free parameter,
and the initial bulk-brane classical action is defined by some 
considerations. The effective bulk-brane equations of motion are analyzed.
The quantum creation of 4d de Sitter or Anti-de Sitter(AdS) brane Universe 
living in 5d AdS space is possible when quantum bulk and (or) brane
matter is taken into account. The consideration of only conformal field 
theory (CFT) living on the brane admits the full analytical treatment.
 Then bulk gravitational
 Casimir effect leads to deformation of 5d AdS space shape as well as of 
shape of spherical or hyperbolic branes. The generalization of above 
picture for the dominant bulk quantum gravity naturally represents such  
scenario as self-consistent warped compactification within AdS/CFT set-up.

\newpage

\section{Introduction}

The idea that observable Universe represents the brane
where 4d gravity is trapped \cite{RS} embedded in 
higher dimensional bulk space attracted enormous attention.
The study of cosmological aspects of such brane-worlds \cite{CH,cosm}
(and refs. therein) indicates towards the possibility to
construct the inflationary brane Universe \cite{CH}. Some 
observational manifestations of bulk matter fields may occur on 
the brane level \cite{exp}. As brane-world Universe is naturally 
realized in more than four dimensional bulk space it could be 
related with recent studies of AdS/CFT correspondence \cite{AdS}.
One attempt could be in implementing of RS warped compactification 
within the context of RG flow in AdS/CFT set-up. It may be done in 
the  simplest form in the way suggested in refs.\cite{NOZ,HHR} via 
consideration of quantum CFT living on the brane. 

Indeed, this scenario significally varies from original approach where 
brane tension (brane cosmological constant) is free parameter of theory 
permitting actually the existence of brane solution by its fine-tuning.
In scenario \cite{NOZ,HHR} the action is defined from the beginning and 
surface terms are added in closed manner. The role of surface terms is to 
make the variational procedure to be well-defined and to cancell the 
leading divergences of the action around AdS bulk space. Hence, brane tension
is not free parameter anymore.
However, its role is taken by quantum effects.
Indeed, quantum 
effects of brane CFT induce the conformal anomaly and anomaly 
induced effective action. This 4d anomaly induced (brane) effective 
action should be added to the complete effective action of 
five-dimensional theory. In fact, it gives explicit contribution to 
brane tension permitting to have the consistent curved brane solutions.
 As the result,
the possibility of quantum creation of de Sitter or Anti-de Sitter 
brane living in 5d AdS Universe has been proved 
in refs.\cite{NOZ,HHR,NOm}. The simplest choice for brane 
CFT is maximally supersymmetric (SUSY) Yang-Mills theory.
Thus, conformal anomaly of brane CFT induces the brane effective tension 
which is responsible for de Sitter (or AdS) geometry of brane. 
Note that anomaly induced effective action may be considered as 
kind of gravitational Casimir effect \cite{Birrell} (for a recent 
introduction to Casimir effect, see \cite{kim}).

Developing further the study of warped compactifications with curved 
boundary (inflationary brane) within AdS/CFT correspondence the 
natural question is about the role of quantum bulk effects in 
such scenario. In other words, quantum effects of five-dimensional bulk 
gravity (i.e. bulk gravitational Casimir effect)  should be taken into 
account. When the boundary of AdS space is flat it can be 
done \cite{garriga} in the analogy with the usual calculation
of quantum effective action in Kaluza-Klein multi-dimensional gravity 
\cite{chodos}
(for a review and complete list of references, see \cite{BOS}).
However, for AdS bulk space with curved boundary such quantum 
gravity calculation is much more involved. This will be done elsewhere.

In the present paper, in order to estimate (at least, qualitatively) 
the role of bulk quantum effects to the scenario of 
refs.\cite{NOZ,HHR,NOm} we consider the contribution of 
quantum bulk matter (on the example of scalar) to complete 
five-dimensional  effective action.
Having the structure of bulk effective action for conformal matter 
we discuss brane-world cosmology where bulk and boundary quantum effects 
are taken into account.
The effective bulk-brane equations of motion are derived and their 
solutions are analyzed.
It is shown that due to such effects the quantum creation of de Sitter
or Anti-de Sitter brane living in 5d AdS Universe is possible,
where bulk quantum effects deform the shape of constant curvature brane
(our observable Universe). Note that such Universe occurs when only  
brane quantum effects are included. Hence, bulk quantum effects modify 
the geometrical configuration which is recovered in their absence. 
It is interesting that above 
quantum creation is possible  
only due to bulk matter quantum effects,i.e. when no CFT lives on the brane.

\def\ii{\'{\i}}
\def\bi{\bigskip}
\def\noi{\noindent}
\def\be{\begin{equation}}
\def\en{\end{equation}}
\def\bq{\begin{eqnarray}}
\def\eq{\end{eqnarray}}

\section{Gravitational Casimir effect for bulk scalar in AdS space}

Let us start from the following lagrangian 
(Euclidean sector) for a conformally invariant 
massless   scalar field with scalar-gravitational coupling 

\be
{\cal L}= \sqrt{g}  \chi  (-\Box+ \xi R ) \chi
\en

where 
$\xi = \frac{{\cal D}-2}{4({\cal D}-1)}$. 
Having in mind the applications 
to Randall-Sundrum scenario one takes ${\cal D}$ to be odd. First of all
one considers the warped metric of the form typical for
warped compactification \cite{RS}:

\be
ds^2 = dy^2 +  \e^{2A(y)} d \Omega_{{\cal D}-1}^2 
= \e^{2A(z)} \left[ dz^2 +
d \Omega_{{\cal D}-1}^2 \right]\,, 
\label{RS}
\en
where $ d \Omega_{{\cal D}-1}^2 $ corresponds to a ${\cal D}$-1-dimensional 
constant curvature symmetric space M$_{{\cal D}-1}$, 
namely R$_{{\cal D}-1}$, S$_{{\cal D}-1}$ and H$_{{\cal D}-1}$, 
the Euclidean space, the sphere and the hyperbolic space respectively. 
Furthermore, the  warp  factor is given by
\be
\e^{A(z)}=\frac{l}{z}\,,
\en
$l$ being  related to the cosmological constant. We may put $l=1$ and later by
dimensional analysis recover it. 
Furthermore, in the following, let us consider ${\cal D}=5$. It may be 
convenient to 
make the conformal transformation for the metric 
\be
(ds^2)'=z^2 ds^2
\en
and for the  scalar field  $\chi'= z^{-3/2}
\chi$. Then, for rescaled scalar $\chi'$, one gets the  
Lagrangian

\be
{\cal L} = \chi' (-\Box'+ \xi R_{(4)} ) \chi'\,,
\en
where $R_{(4)}$ is the constant scalar curvature of M$_4$ and 
$\Box'$ is the Laplace operator in the product space 
${\rm R} \times {\rm M}_4$, whose domain is subject to suitable
boundary conditions at $z=l$ and $z=L$, induced by the orbifold 
nature of the
space-time we have started with. 

Now,  one should 
calculate the one-loop 
effective potential, {\it i.e.} the effective action divided by the whole 
volume
\be
V=\frac{1}{2 L {\rm Vol}
({\rm M}_4)} \log \det  \left(L_5 /\mu^2 \right)\,,
\en
where 
\be
L_5=-\partial_z^2   - \Box_{(4)} + \xi R_{(4)}=L_1+L_4  \,,  
\en
on    ${\rm R} \times {\rm M}_4$  limited by two
branes subject to boundary conditions. At the end, 
if $L$ is the branes separation, we  
will  take the limit $ L$ goes to infinity.

Since, one is dealing with a product space, the heat-kernel for 
$L_5$ is given by
\be
K_t(L_5)=K_t(L_1)K_t(L_4)\,,\
\en
where 
\be
K_t(L_1)=\sum_n \e^{-t\lambda^2_n} 
\en
 is the heat-kernel related to the one-dimensional operator
$L_1= -\partial_z^2  $ whose domain contains the conformally transformed 
orbifold 
boundary conditions, $\lambda_n$ the associated eigenvalues and  
\be
K_t(L_4)=\sum_{\alpha} \e^{-t\lambda^2_\alpha}  
\en
is the heat-kernel of the Laplace-like operator $L_4$ on M$_4$, 
$\lambda_\alpha=\mu_\alpha+\xi R_{(4)}$,  
$\mu_\alpha$ being the eigenvalues of $-\Box_{(4)}$. 

If we make use of the zeta-function regularization, one needs the analytical 
continuation of the zeta-function
\be
\zeta(s|L_5)=\sum_{\alpha}\sum_{n=1}^\infty 
(\lambda_n+\lambda_\alpha^2)^{-s}\,. 
\en

Generally speaking, we do not know explicitly the spectrum of $L_1$. 
However, it is known the short $t$ asymptotics, which is given by
\be
K_t(L_1)=\sum_{r=0}^\infty K_r(L_1)t^{\frac{r-1}{2}}
=\frac{L}{2\sqrt{\pi t}}+
K_1(L_1)+O(t^{1/2})+O\left(\frac{1}{L}\right)
+O(\e^{-\frac{L^2}{t}})\,.  
\en
Here $L$ is the brane separation, the leading term is the Weyl term and the 
next term is the first non-trivial boundary term, which, for 
dimensional reasons, is  a numerical constant. Since there is no potential 
term, the other boundary terms are, again for dimensional reasons, of order $
O(1/L)$. Thus,  if $L$ goes to infinity, the above 
asymptotics becomes almost exact. As a result, we have 
\be
\zeta(s|L_5)=\frac{L\Gamma\left(s-\frac{1}{2}\right)}
{2\Gamma(s)}
\zeta\left(\left.s-\frac{1}{2}\right|L_4\right)+K_1(L_1)\zeta(s|L_4)+O\left(\frac{1}{L}
\right)\,,
\en 
where $\zeta(s|L_4)$ is the zeta-function associated with the Laplace-like 
operator on M$_4$. 

It should be noted that 
\be
\zeta(0|L_5)=K_1(L_1)\zeta(0|L_4)\,.
\en
Thus, there exists also a non trivial 
contribution
coming from the Jacobian related to the conformal transformation 
we have performed. We are neglecting for the moment this 
contribution as we argue later on
that it is negligible. 

As a consequence, in the large $L$ limit, 
the effective potential  reads
\bq
V&=&-\frac{1}{2L {\rm Vol}({\rm M}_4)}
[\zeta'(0|L_5)+\ln \mu^2 \zeta(0|L_5)] \nonumber \\ 
&=&\frac{\sqrt \pi}{2 {\rm Vol}({\rm M}_4)} 
\zeta\left(\left.-\frac{1}{2}\right|L_4\right)
+ O\left(\frac{1}{L}\right)
\,,
\eq

In this limit, the effective potential, reduces to the Casimir 
energy (vacuum energy) related to the Laplace-like  operator on M$_4$. This 
Casimir energy 
 has been calculated in 
several places (see for example, \cite{camporesi90,elib,bytsenko96,dolan} and 
references 
therein). We recall 
the corresponding results. 

First, in the case of flat brane R$_4$, the Casimir energy goes like 
$(O(1/L^4)$, 
thus it is negligible. 

For the spherical brane S$_4$ (with radius ${\cal R}$), the starting point is
\be
\zeta(s|L_4)=g(s){\cal R}^{2s}
\en
where
\be
g(s)=\frac{1}{6}\sum_{l=1}^\infty (l+1)(l+2)(2l+3)
\left(l^2+3l+\frac{9}{4}\right)^{-s}\,,
\en
the analytical continuation can be easily done, due to conformal coupling in 
5 dimensions and the result is 
\be
\zeta(s|L_4)=\frac{{\cal R}^{2s}}{3}\left[ 
\zeta_H\left(2s-3,\frac{3}{2}\right)
-\frac{1}{4}\zeta_H\left(2s-1,\frac{3}{2}\right) \right]\,,
\en
where $\zeta_H(s,a)$ is the Hurwitz zeta-function. 
Thus,
\be
\zeta\left(\left. -\frac{1}{2}\right|L_4\right)
=\frac{1}{3{\cal R}}\left[\zeta_H\left(
-4,\frac{3}{2}\right)
-\frac{1}{4}\zeta_H\left(-1,\frac{3}{2}\right) \right]\,.
\en
Making use of
\be
\zeta_H(-m,a)=-\frac{B_{m+1}(a)}{m+1}\,,
\en
where $B_n(x)$ is a Bernoulli polynomial, one gets $\zeta(-\frac{1}{2}|L_4)=0$.
In this case $V=0$.
This is also consistent with the result reported in ref. \cite{dolan}.
Note that taking non-conformal coupling constant in the initial 
Lagrangian 
changes qualitatively this result, then potential will not be zero anymore.
It could be also non-zero for another matter fields.

In the hyperbolic brane H$_4$, one has \cite{bytsenko96} 
\be
\frac{\zeta(s|L_4)}{{\rm Vol}({\rm H}_4)}=
-\frac{{\cal R}^{2s-4}}{4 \pi^2}\int_0^\infty 
\lambda^{-2s+1}
\frac{(\lambda^{2}+\frac{1}{4})}{\e^{2\pi \lambda}+1}
d\lambda\,.
\en
Thus, the effective potential reads
\be
V=-\frac{1}{8 \pi^{3/2}{\cal R}^5}\int_0^\infty 
\frac{(\lambda^{2}+\frac{1}{4})}{\e^{2\pi \lambda}+1}
d\lambda\,.
\en
The integral can be easily evaluated. One has
\be
V=-\frac{1}{64 \pi^{5/2}{\cal R}^5} \left[\ln 2
+\frac{3}{4 \pi^2}\zeta_R(3)\right]\,.
\en
Note that back conformal transformation of this potential should be done,
to recover the potential in the original AdS space. This transformation
introduces the factor $\e^{-5A}$ in the above expression.

With regard to the Jacobian factor, one may  introduce the interpolating 
\be
(ds^2_q)=z^{2q} ds^2
\en
with $q$ a real parameter such that if $q=0$, then $(ds^2_0)= ds^2$ and if
$q=1$, then $(ds^2_1)= (ds^2)'$. Then (see, for example, \cite{bytsenko96})
\be
\ln J(g,g')=\int_0^1 eq \int_l^L dz z^{5(q-1)}
\int \sqrt{g_4}d^4x \zeta(0|L_5(q))(z,x)\,,
\en
with
\be
L_5(q)=-\Box_{(5)}(q)+ \xi R_{(5)}(q)
\en
the conformal scalar operator in the interpolating metric and 
$\zeta(0|L_5(q))(z,x)=\frac{K_5}{(4\pi)^{5/2}}$ is the local zeta-function.
In a 5-dimensional manifold without boundary, $K_5=0$. In our case, 
however, we have orbifold boundary conditions and the coefficient $K_5$ is not 
vanishing. It has  been recently computed for Robin boundary condition in 
\cite{branson98}. Unfortunately the expression given in the above reference is
difficult to use in our case. 

We may assume that $K_5(q)$ is a linear combination of terms 
\be
\sum_l a_l q^l+b_l(1-q)^l\,. 
\en
Then, we have 
\be
\ln J(g,g')=\sum_l \int_0^1 dq \int_l^L dz z^{5(q-1)} (a_l q^l+b_l(1-q)^l) \,.
\en
This integral gives  contributions proportional to $li(L)$, where $li(z)$ 
is the 
logarithmic integral.  For $L $ very large, the logarithmic integral  
diverges as
$ O(\frac{L}{\ln L})$. Then, it gives a contribution of order 
$ O(\frac{1}{\ln L})$ to the effective potential.

Thus, we presented the explicit example 
of evaluation of gravitational Casimir effect (effective potential)
for conformal bulk scalar on five-dimensional AdS space 
with 4-dimensional sphere or hyperboloid as a boundary.
This calculation gives us an idea about the structure 
of effective action due to bulk matter quantum effects.


\section{Quantum creation of de Sitter (Anti-de Sitter) 
brane-world Universe}

We consider the spacetime whose boundary is four-dimensional 
sphere S$_4$, which can be identified with a D3-brane or
four-dimensional hyperboloid H$_4$. 
The bulk part is given by 5 dimensional 
Euclidean Anti-de Sitter space $\AdS5$ 
\be
\label{AdS5i}
ds^2_\AdS5=dy^2 + \sinh^2 {y \over l}d\Omega^2_4\ .
\ee
Here $d\Omega^2_4$ is given by the metric of S$_4$ or H$_4$ 
with unit radius. One also assumes the boundary (brane) 
lies at $y=y_0$ 
and the bulk space is given by gluing two regions 
given by $0\leq y < y_0$ (see\cite{HHR} for more details.)

We start with the action $S$ which is the sum of 
the Einstein-Hilbert action $\SEH$, the Gibbons-Hawking 
surface term $\SGH$ \cite{GH},  the surface counter term $S_1$ 
and the trace anomaly induced action $W$\footnote{For the 
introduction to anomaly induced effective action in curved 
space-time (with torsion), see section 5.5 in \cite{BOS}.}: 
\bea
\label{Stotal}
S&=&\SEH + \SGH + 2 S_1 + W \\
\label{SEHi}
\SEH&=&{1 \over 16\pi G}\int d^5 x \sqrt{\gfv}\left(R_{(5)} 
+ {12 \over l^2}\right) \\
\label{GHi}
\SGH&=&{1 \over 8\pi G}\int d^4 x \sqrt{\gfr}\nabla_\mu n^\mu \\
\label{S1}
S_1&=& -{3 \over 8\pi G}\int d^4 x \sqrt{\gfr} \\
\label{W}
W&=& b \int d^4x \sqrt{\widetilde g}\widetilde F A \nn
&& + b' \int d^4x \sqrt{\widetilde g}
\left\{A \left[2{\widetilde\Box}^2 
+\widetilde R_{\mu\nu}\widetilde\nabla_\mu\widetilde\nabla_\nu 
 - {4 \over 3}\widetilde R \widetilde\Box^2 
+ {2 \over 3}(\widetilde\nabla^\mu \widetilde R)\widetilde\nabla_\mu
\right]A \right. \nn
&& \left. + \left(\widetilde G - {2 \over 3}\widetilde\Box 
\widetilde R
\right)A \right\} \nn
&& -{1 \over 12}\left\{b''+ {2 \over 3}(b + b')\right\}
\int d^4x \sqrt{\widetilde g} \left[ \widetilde R 
- 6\widetilde\Box A 
 - 6 (\widetilde\nabla_\mu A)(\widetilde \nabla^\mu A)
\right]^2 \ .
\eea 
Here the quantities in the  5 dimensional bulk spacetime are 
specified by the suffices $_{(5)}$ and those in the boundary 4 
dimensional spacetime are by $_{(4)}$. 
The factor $2$ in front of $S_1$ in (\ref{Stotal}) is coming from 
that we have two bulk regions which 
are connected with each other by the brane. 
In (\ref{GHi}), $n^\mu$ is 
the unit vector normal to the boundary. In (\ref{W}), one chooses 
the 4 dimensional boundary metric as 
\be
\label{tildeg}
\gfr_{\mu\nu}=\e^{2A}\tilde g_{\mu\nu}
\ee 
and we specify the 
quantities with $\tilde g_{\mu\nu}$ by using $\tilde{\ }$. 
$G$ ($\tilde G$) and $F$ ($\tilde F$) are the Gauss-Bonnet
invariant and the square of the Weyl tensor
\footnote{We use the following curvature conventions:
\begin{eqnarray*}
R&=&g^{\mu\nu}R_{\mu\nu} \\
R_{\mu\nu}&=& R^\lambda_{\ \mu\lambda\nu} \\
R^\lambda_{\ \mu\rho\nu}&=&
-\Gamma^\lambda_{\mu\rho,\nu}
+ \Gamma^\lambda_{\mu\nu,\rho}
- \Gamma^\eta_{\mu\rho}\Gamma^\lambda_{\nu\eta}
+ \Gamma^\eta_{\mu\nu}\Gamma^\lambda_{\rho\eta} \\
\Gamma^\eta_{\mu\lambda}&=&{1 \over 2}g^{\eta\nu}\left(
g_{\mu\nu,\lambda} + g_{\lambda\nu,\mu} - g_{\mu\lambda,\nu} 
\right)\ .
\end{eqnarray*}}
\bea
\label{GF}
G&=&R^2 -4 R_{ij}R^{ij}
+ R_{ijkl}R^{ijkl} \nn
F&=&{1 \over 3}R^2 -2 R_{ij}R^{ij}
+ R_{ijkl}R^{ijkl} \ ,
\eea
In the effective action (\ref{W}), with $N$ scalar, $N_{1/2}$ 
spinor, $N_1$ vector fields, $N_2$ ($=0$ or $1$) gravitons 
and $N_{\rm HD}$ higher 
derivative conformal scalars, $b$, $b'$ and $b''$ are 
\bea
\label{bs}
b&=&{N +6N_{1/2}+12N_1 + 611 N_2 - 8N_{\rm HD} \over 120(4\pi)^2}\nn 
b'&=&-{N+11N_{1/2}+62N_1 + 1411 N_2 -28 N_{\rm HD} \over 360(4\pi)^2}\ , 
\nn 
b''&=&0\ .
\eea
As usually, $b''$ may be changed by the finite 
renormalization of local counterterm in gravitational 
effective action. As we shall see later, the term proportional 
to $\left\{b''+ {2 \over 3}(b + b')\right\}$ in (\ref{W}), and 
therefore $b''$, does not contribute to the equations of motion. 
For ${\cal N}=4$ $SU(N)$ super Yang-Mills theory 
$b=-b'={N^2 -1 \over 4(4\pi )^2}$. 
As one can see until this point the discussion repeats the one
presented in ref.\cite{NOm} where more detail may be found. It is 
interesting to 
note that the contribution from brane quantum gravity may be taken into 
account via the correspondent coefficient in above equation.

We should also note that $W$ in (\ref{W}) is defined up to 
conformally invariant functional, which cannot be determined 
from only the conformal anomaly. The conformally flat space is 
a pleasant exclusion where anomaly induced effective action 
is defined uniquely. However, one can argue that such 
conformally invariant functional gives 
next to leading contribution as mass parameter of regularization 
may be adjusted to be arbitrary small (or large).

The metric of $\S4$ with the unit radius is given by
\be
\label{S4metric1}
d\Omega^2_4= d \chi^2 + \sin^2 \chi d\Omega^2_3\ .
\ee
Here $d\Omega^2_3$ is the metric of 3 dimensional 
unit sphere. If we change the coordinate $\chi$ to 
$\sigma$ by 
\be
\label{S4chng}
\sin\chi = \pm {1 \over \cosh \sigma} \ , 
\ee
one obtains
\be
\label{S4metric2}
d\Omega^2_4= {1 \over \cosh^2 \sigma}\left(d \sigma^2 
+ d\Omega^2_3\right)\ .
\ee
On the other hand, the metric of the 4 dimensional flat 
Euclidean space is 
\be
\label{E4metric}
ds_{\rm 4E}^2= d\rho^2 + \rho^2 d\Omega^2_3\ .
\ee
Then by changing the coordinate as 
\be
\label{E4chng}
\rho=\e^\sigma\ , 
\ee
one gets
\be
\label{E4metric2}
ds_{\rm 4E}^2= \e^{2\sigma}\left(d\sigma^2 + d\Omega^2_3\right)\ .
\ee
For the 4 dimensional hyperboloid with the unit radius, 
the metric is 
\be
\label{H4metric1}
ds_{\rm H4}^2= d \chi^2 + \sinh^2 \chi d\Omega^2_3\ .
\ee
Changing the coordinate $\chi$ to $\sigma$  
\be
\label{H4chng}
\sinh\chi = {1 \over \sinh \sigma} \ , 
\ee
one finds
\be
\label{H4metric2}
ds_{\rm H4}^2 = {1 \over \sinh^2 \sigma}\left(d \sigma^2 
+ d\Omega^2_3\right)\ .
\ee

Motivated by (\ref{AdS5i}), (\ref{S4metric2}), 
(\ref{E4metric2}) and (\ref{H4metric2}), one assumes 
the metric of 5 dimensional space time as follows:
\be
\label{metric1}
ds^2=dy^2 + \e^{2A(y,\sigma)}\tilde g_{\mu\nu}dx^\mu dx^\nu\ ,
\quad \tilde g_{\mu\nu}dx^\mu dx^\nu\equiv l^2\left(d \sigma^2 
+ d\Omega^2_3\right)
\ee
and we identify $A$ and $\tilde g$ in (\ref{metric1}) with those in 
(\ref{tildeg}). Then one finds $\tilde F=\tilde G=0$, 
$\tilde R={6 \over l^2}$ etc. 
Due to Eq. (\ref{metric1}), the actions in (\ref{SEHi}), 
(\ref{GHi}), (\ref{S1}), and (\ref{W}) have the following forms:
\bea
\label{SEHii}
\SEH&=& {l^4 V_3 \over 16\pi G}\int dy d\sigma \left\{\left( -8 
\partial_y^2 A - 20 (\partial_y A)^2\right)\e^{4A} \right. \nn
&& \left. +\left(-6\partial_\sigma^2 A - 6 (\partial_\sigma A)^2 
+ 6 \right)\e^{2A} + {12 \over l^2} \e^{4A}\right\} \\
\label{GHii}
\SGH&=& {3l^4 V_3 \over 8\pi G}\int d\sigma \e^{4A} 
\partial_y A \\
\label{S1ii}
S_1&=& - {3l^3 V_3 \over 8\pi G}\int d\sigma \e^{4A} \\
\label{Wii}
W&=& V_3 \int d\sigma \left[b'A\left(2\partial_\sigma^4 A
 - 8 \partial_\sigma^2 A \right) \right. \nn
&&\left. - 2(b + b')\left(1 - \partial_\sigma^2 A 
 - (\partial_\sigma A)^2 \right)^2 \right]\ .
\eea
Here $V_3=\int d\Omega_3$ is the volume or area of 
the unit 3 sphere.

As it follows from the discussion in the previous section 
there is also gravitational Casimir contribution due
to bulk quantum fields. As one sees on the example of 
bulk scalar it has typically the following form $S_{\rm Csmr}$ 
\be
\label{SCsmr}
S_{\rm Csmr}={cV_3 \over {\cal R}^5}\int dy 
d\sigma \e^{-A} 
\ee
Note that role of (effective) radius of 4d constant curvature space (after 
back conformal
transformation as in previous section) is played by ${\cal R}\e^{A}$.
Here $c$ is some coefficient whose value and sign depend 
on the type of bulk field (scalar, spinor, vector, graviton, ...) 
and on parameters of bulk theory (mass, scalar-gravitational 
coupling constant, etc). In the previous section we found this 
coefficient for conformal scalar. In the following discussion
it is more convenient to consider this coefficient to be some 
parameter of the theory. Then, the results are quite common and 
may be applied to arbitrary quantum bulk theory. We also suppose 
that there are no background
bulk fields in the theory (except of bulk gravitational field).

Adding quantum bulk contribution
to the action $S$ in (\ref{Stotal}) one can regard 
\be
\label{StotalB}
S_{\rm total}=S+S_{\rm Csmr}
\ee
as the total action. In (\ref{SCsmr}), ${\cal R}$ is 
the radius of S$_4$ or H$_4$. 

In the bulk, one obtains the following equation of motion 
from $\SEH + S_{\rm Csmr}$ by the variation over $A$:
\bea
\label{eq1}
0&=& \left(-24 \partial_y^2 A - 48 (\partial_y A)^2 
+ {48 \over l^2}\right)\e^{4A} \nn
&& + {1 \over l^2}\left(-12 \partial_\sigma^2 A 
- 12 (\partial_\sigma A)^2 + 12\right)\e^{2A}
+ {16\pi G c \over {\cal R}^5 }\e^{-A}\ .
\eea
First, one can consider a special solution of the bulk equation 
(\ref{eq1}). If one assumes that $A$ does not depend on $\sigma$, 
Eq.(\ref{eq1}) has the following form:
\be
\label{eq1b}
0= \left(-24 \partial_y^2 A - 48 (\partial_y A)^2 
+ {48 \over l^2}\right)\e^{4A} 
+ {16\pi G c \over {\cal R}^5 }\e^{-A}\ .
\ee
Eq.(\ref{eq1b}) has the following integral:
\be
\label{Int}
E=-{1 \over 4}\left({d \left(\e^{2\tilde A}\right) \over dy}
\right)^2 + {1 \over l^2}\e^{4\tilde A} 
+ {1 \over 2l^2}\e^{2\tilde A}
- {4\pi G c \over 3 {\cal R}^5} \e^{-\tilde A}\ .
\ee
Then if we assume ${d \left(\e^{2\tilde A}\right) 
\over dy}>0$, 
we find the following solution in the bulk
\be
\label{sly}
y={1 \over 2}\int dQ \left(-E+{Q^2 \over l^2} 
+ {Q \over 2l^2}- {4\pi G c \over 3 {\cal R}^5} 
Q^{-{1 \over 2}}\right)^{-{1 \over 2}}\ .
\ee

This represents an example of self-consistent warped compactification.
The analysis of such solution shows that for vanishing bulk cosmological 
constant it goes away from AdS space. This indicates that bulk 
Casimir effect acts against of warped compactification.

Let us discuss the solution in the situation when scale factor depends 
on both coordinates:$y$,$\sigma$.
One can find the solution of (\ref{eq1}) as an expansion 
with respect to $\e^{-{y \over l}}$ by assuming that ${y \over l}$ 
is large:
\be
\label{S4sl}
\e^A={\sinh {y \over l} \over \cosh \sigma}
- {32\pi Gcl^3 \over 15 {\cal R}^5}\cosh^4\sigma 
\e^{-{4y \over l}} + {\cal O}\left(\e^{-{5y \over l}}\right)
\ee
for the perturbation from the solution where the brane 
is S$_4$ and 
\be
\label{H4sl}
\e^A={\cosh {y \over l} \over \sinh \sigma}
- {32\pi Gcl^3 \over 15 {\cal R}^5}\sinh^4\sigma 
\e^{-{4y \over l}} + {\cal O}\left(\e^{-{5y \over l}}\right)
\ee
for the perturbation from H$_4$ brane solution.

On the brane at the boundary, 
one gets the following equation: 
\bea
\label{eq2}
0&=&{48 l^4 \over 16\pi G}\left(\partial_y A - {1 \over l}
\right)\e^{4A}
+b'\left(4\partial_\sigma^4 A - 16 \partial_\sigma^2 A\right) \nn
&& - 4(b+b')\left(\partial_\sigma^4 A + 2 \partial_\sigma^2 A 
 - 6 (\partial_\sigma A)^2\partial_\sigma^2 A \right)\ .
\eea
We should note that the contributions from $\SEH$ and $\SGH$ are 
twice from the naive values since we have two bulk regions which 
are connected with each other by the brane. 
Substituting the solutions (\ref{S4sl}) and (\ref{H4sl}) 
into (\ref{eq2}), we find
\be
\label{S4slbr2}
0\sim{1 \over \pi G}\left({1 \over {\cal R}}
\sqrt{1 + {{\cal R}^2 \over l^2}}
 + {64\pi G l^7 c \over 3 {\cal R}^{10}}\cosh^5\sigma
 - {1 \over l}\right){\cal R}^4 + 8b'\ .
\ee
for S$_4$ brane and 
\be
\label{H4slbr2}
0\sim{1 \over \pi G}\left({1 \over {\cal R}}
\sqrt{-1 + {{\cal R}^2 \over l^2}}
 + {64\pi G l^7 c \over 3 {\cal R}^{10}}\sinh^5\sigma
 - {1 \over l}\right){\cal R}^4 + 8b'\ .
\ee
for H$_4$ brane.
Here the radius ${\cal R}$ of S$_4$ or H$_4$ is related 
with $A(y_0)$, if we assume the brane lies at $y=y_0$, by
\be
\label{tldR}
\tilde R = l\e^{\tilde A(y_0)}\ .
\ee
In Eqs.(\ref{S4slbr2}) and (\ref{H4slbr2}), only the leading terms 
with respect to $1/{\cal R}$ are kept in the ones coming from 
$S_{\rm Csmr}$ (the terms including $c$). 
When $c=0$, the previous result in \cite{HHR,NOm} is reproduced. 
Eqs.(\ref{S4slbr2}) and (\ref{H4slbr2}) tell that 
the Casimir force deforms the shape of S$_4$ or H$_4$ since 
${\cal R}$ becomes $\sigma$ dependent. The effect 
becomes larger for large $\sigma$. In case of 
S$_4$ brane, the effect becomes large if the distance from the 
equator becomes large since $\sigma$ is related to the angle 
coordinate $\chi$ by (\ref{S4chng}). Especially at north and 
south poles ($\chi=0$, $\pi$), $\cosh\sigma$ diverges then 
${\cal R}$ should vanish as in Fig.1. 
Of course, the perturbation would be invalid when $\cosh\sigma$ 
is large. Thus, we demonstrated that bulk quantum effects 
do not destroy the quantum creation of de Sitter (inflationary) 
or Anti-de Sitter brane-world Universe. Of course, analytical 
continuation of 4d sphere to Lorentzian signature is supposed 
which leads to ever expanding inflationary brane-world Universe.
However, as we see the bulk quantum effects change the effective 
radius of 4d sphere (or 4d hyperboloid).

When $c=0$, the solution can exist when $b'<0$ for S$_4$ 
brane (in this case it is qualitatively similar to quite well-known
anomaly driven inflation of refs.\cite{SMM}) and $b'>0$ for H$_4$. 
For S$_4$ brane, if $b'<0$, the effect 
of Casimir force makes the radius smaller (larger) if $c>0$ ($c<0$).
For H$_4$ brane, from Eq.(\ref{H4slbr2})  for small 
${\cal R}$ it behaves as 
\be
\label{H4slbr3}
0\sim {64 l^7 c \over 3 {\cal R}^{10}}\sinh^5\sigma
+ 8b'\ .
\ee
Then one would have solution even if $b'<0$. (We should note 
$\sigma$ is restricted to be positive). Of course, 
one cannot do any quantitative conclusion  since it is 
assumed ${\cal R}$ is large when deriving (\ref{H4slbr2}).

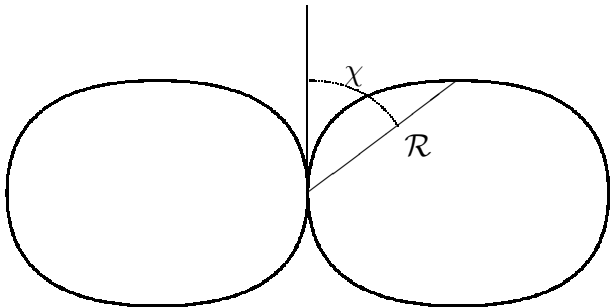
\begin{figure}
\setlength{\unitlength}{1mm}
\begin{picture}(80,40)
\put(60,20){\line(0,1){25}}
\put(60,20){\line(4,3){20}}
\qbezier[40](60,35)(68,35)(72,29)
\put(73,25){${\cal R}$}
\put(65,35){$\chi$}
\thicklines
\qbezier[100](60,20)(60,35)(80,35)
\qbezier[100](80,35)(100,35)(100,20)
\qbezier[100](60,20)(60,5)(80,5)
\qbezier[100](80,5)(100,5)(100,20)
\qbezier[100](60,20)(60,35)(40,35)
\qbezier[100](40,35)(20,35)(20,20)
\qbezier[100](60,20)(60,5)(40,5)
\qbezier[100](40,5)(20,5)(20,20)
\end{picture}
\caption{\label{Fig1} The rough sketch of the 
shape of cross section of the S$_4$brane when $c>0$.}
\end{figure}

We now compare the above obtained results with the case 
of no quantum correction $W$ (no matter) on the brane,
i.e. when bulk quantum effects are leading.  Putting 
$b'=0$ in (\ref{S4slbr2}) and (\ref{H4slbr2}), one gets
\be
\label{S4slbrC}
{\cal R}^8\sim -{128\pi G l^6 c \over 8}\cosh^5\sigma
\ee
for S$_4$ brane and 
\be
\label{H4slbrC}
{\cal R}^8\sim {128\pi G l^6 c \over 8}\sinh^5\sigma
\ee
for H$_4$ brane. Here we only consider the leading term with 
respect to $c$, which corresponds to large ${\cal R}$ 
approximation.
In case of S$_4$, there is no real solution for negative $b'$ 
but there appears a solution for negative $b'$ in case of H$_4$, 
where there is no solution without 
Casimir term $S_{\rm Csmr}$ in (\ref{SCsmr}). 
Thus, we demonstrated that bulk quantum effects do not violate
(in some cases ,even support) the quantum creation of de Sitter 
or Anti-de Sitter brane living in d5 AdS world.

\section{Discussion}

In summary, we compared the role of bulk matter and brane matter 
quantum effects in the realization of brane-world Universe 
with constant curvature 4d brane (de Sitter or Anti-de Sitter).
In such Universe the bulk represents five-dimensional AdS space while
observable four-dimensional Universe is the boundary (brane) of such 
five-dimensional space.
The brane matter quantum effects may be included in the universal form,
via the corresponding anomaly induced effective action on the brane.
(Actually, they modify the effective brane tension which is fixed on 
classical level). 
In this way, the contribution of any specific conformally invariant 
matter theory (or even of brane quantum gravity) only changes the 
coefficients of the correspondent anomaly induced effective action.
The bulk conformal matter quantum effects are more difficult to calculate.
We made the correspondent evaluation for scalar in order to understand
the qualitative structure of bulk effective action. 

It is shown that quantum creation of inflationary (or hyperbolic) 
brane-world Universe is possible. Such Universe may be induced by only
bulk quantum effects , or by only brane quantum effects. When both
contributions are included the role of bulk quantum effects is in the 
deformation
of shape of 5d AdS Universe as well as of shape of 4d spherical or 
hyperbolic
brane. (These are induced by brane CFT quantum effects). 

There are few possible extensions of the presented results.
As we already mentioned in the introduction, it would be really 
interesting (and more consistent from the AdS/CFT correspondence 
point of view) to investigate the role of bulk quantum gravity 
in such self-consistent warped Randall-Sundrum compactification with curved 
boundary. This is under study currently.

 From another side, one can combine the current scenario for 
quantum induced inflationary brane-world Universe with 
the ones \cite{CH} where classical background matter is presented in the 
bulk and (or) on the brane. In any case, the number of possibilities to
realize brane-world inflation occurs. 
\noindent

{\bf Acknowledgments.} 
We would like to thank A. Bytsenko and K. Milton for helpful
discussions.
The work by SDO has been supported in part by CONACyT (CP, ref.990356 and 
grant 28454E) 
and in part by RFBR. The work by SZ and SDO has been supported in part by 
INFN, Gruppo Collegato di Trento.

\end{document}